\documentclass[epjST]{svjour}
\usepackage{graphics}
\usepackage[latin1]{inputenc}
\usepackage{calc}
\usepackage{setspace}
\usepackage{fixltx2e}
\usepackage{graphicx}
\usepackage{multicol}
\usepackage[normalem]{ulem}
\usepackage[english]{babel}
\usepackage{color}
\usepackage{hyperref}
\usepackage{enumerate}
\usepackage{graphics}
\usepackage[english]{babel}
\usepackage{mdframed,xcolor}
\usepackage{hyperref}
\hypersetup{
colorlinks=true,
citecolor=blue,
linkcolor=blue,
filecolor=blue,
urlcolor=blue}

\begin{document}
\title{Towards a Living Earth Simulator}
\author{Mario Paolucci\inst{1}\fnmsep\thanks{\email{mario.paolucci@istc.cnr.it}} \and Donald Kossman \inst{2} \and Rosaria Conte\inst{1} \and Paul Lukowicz\inst{3} \and Panos Argyrakis\inst{4} \and Ann Blandford\inst{5} \and Giulia Bonelli\inst{1} \and Stuart Anderson\inst{6} \and Sara de Freitas\inst{7} \and Bruce Edmonds\inst{8} \and Nigel Gilbert\inst{9} \and Markus Gross\inst{2} \and J\"{o}rn Kohlhammer\inst{10} \and Petros Koumoutsakos\inst{2} \and Andreas Krause\inst{2}  \and Bj\"{o}rn-Ola Linn\'{e}r\inst{12} \and Philipp Slusallek\inst{3} \and Olga Sorkine\inst{2} \and Robert W. Sumner\inst{13}\and Dirk Helbing\inst{2}}
\institute{ISTC-CNR, Rome, Italy \and ETH Zurich, Switzerland \and DFKI, Germany \and Aristotle University of Thessaloniki, Greece \and UCL, UK \and University of Edinburgh, UK \and Coventry University, UK \and CPM, Manchester Metropolitan University, UK \and CRESS, University of Surrey, UK \and Fraunhofer Institute for Computer Graphics Research (IGD), Germany  \and Link\"{o}ping University, Sweden \and Disney Research Zurich, Switzerland}
\abstract{
The Living Earth Simulator (LES) is one of the core components of the FuturICT architecture. It will work as a federation of methods, tools, techniques and facilities supporting all of the FuturICT simulation-related activities to allow and encourage interactive exploration and understanding of societal issues. Society-relevant problems will be targeted by leaning on approaches based on complex systems theories and  data science in tight interaction with the other components of FuturICT. The LES will evaluate and provide answers  to real-world questions  by taking into account multiple scenarios. It will build on present approaches such as agent-based simulation and modeling, multiscale modelling, statistical inference, and data mining, moving beyond disciplinary borders to achieve a new perspective on complex social systems.
} 
\maketitle

\tableofcontents

\section{Vision}
 \label{Vision}

Due to the convergence of technological, informational and societal advances, humanity is ready today to make a significant step forward in the direction of the understanding, forecasting and managing of collective activities.

Old and novel problems affect the humanity at local and global scales. They are wicked societal problems, difficult to predict (financial crises,  war), hard to solve (low compliance, poverty), often interconnected and interdependent. These problems typically require an interdisciplinary and cross-methodological treatment (price volatility and political instability), showing non-linear dynamics (rebellions and social conflicts), subject to social contagion (criminality), and likely to raise serious social alarm (epidemics). To deal successfully with such questions in the new interconnected world we are immersed in, we need to lean on Big Data (massive data from heterogeneous sources) collection and processing. A large effort at the frontiers of scientific investigation is thus required. Data, however, are just a part of a larger picture and cannot be used without theories making sense of it. This is why the FuturICT project  will rely on the confluence of disciplines, in particular ICT, complexity science and the social sciences \cite{N1}. It is the social sciences that will provide us with the  unconventional and proud insights about the nature of social behavior. We emphasize the need of a theoretical level as the only one that could really connect society with the mind, and reveals the implicit semantic assumptions that are unavoidably present in data collections.               	                  	                  	
\par

In the recent years, social scientists have started to organize and classify the number, variety, and severity of criticality, if not pathologies and failures, recurring in complex social systems, which involve questions such as \cite{N27},\cite{Helbing and Balietti 2010}:

\begin{itemize}
\item How to understand creativity and innovation?
\item 	How can the formation of social norms and conventions, social roles and socialization, conformity and integration be understood?
\item  How do language and culture evolve?
\item How to comprehend the formation of group identity and group dynamics?
\item How do social differentiation, specialization, inequality and segregation come about?
\item How to model deviance and crime, conflicts, violence, and wars?
\end{itemize}

Another similar list, ranging from how can we persuade people to look after their health, to rather vague questions such as how humanity can increase its {``}collective wisdom''  has been reviewed in \cite{Giles 2011}.

To effectively tackle such questions, which in turn amounts to understanding society at a new level, the necessity to scale up the scientific ambition is evident. The sciences of society and the natural and technological sciences have achieved extraordinary results in the previous century, however they have followed separate lines of development. A joint effort is required today, that will not simply imply that a science of society will be an application of natural science to society. Besides scaling up, we need to change the way society is observed, as argued in \cite{N14}.
Indeed, things are starting to change, also thanks to the transition to socio-technical systems, and the role that participatory ICT systems play in them. Socio-technical systems are complex systems of variable scale, characterized by strong technological components; they are \emph{hybrid} systems because they include both human agents and semi-autonomous automated systems of different degree of  complexity, with which humans may directly interact and by which they more easily interact with one another. Science and technology are thus becoming major actors in the reality they investigate. This new form of interaction makes us anticipate the awakening of a new kind of society where the action of humans, interweaved by ICT systems, could bring about a self-aware society such as \cite{Nowak et al. 2006}.
One of the main functions of the self-awareness is the ability to forecast future events and to plan, reactively and pro actively, towards the fulfillment of the own goals. To build collective self-awareness  means to endow society with forecasting capabilities, even if only short-term. For this, we must leverage ICT tools and complexity science methods that will increase and connect the innate estimation and forecasting capabilities of humans in the sense of crowd wisdom. In this contribution, ideas underlying the design of an innovative ICT system aimed at societal forecasting are discussed.
Particular attention must be paid here to the meaning of forecasting. For a detailed discussion of the possibilities and the limitations of forecasting see \cite{Helbing et al. 2011}. In complex social systems, the cases in which exact predictions can be made are rare. The possible futures of society are in number so large that we can't even begin to enumerate them; they are reflexive and emerge in the sense that every change is able to modify the landscape of its own selection. We can`t predict the result of the game we play because we keep changing the rules of the game. Evolutionary theory,  as stated in \cite{Kauffman2011}, is not `epistemologically closed' because it  cannot list possible futures, and thus it can`t predict. The same line of reasoning holds for predictions regarding the future of society.
Long-term forecasts are simply not a goal of FuturICT. Instead, we intend to understand causal interdependencies, parameter dependencies and to recognize clues and signals indicating the possibility that something significant will happen or not, but not exactly \emph{when} this will happen. In actual social systems, random, chaotic, intrinsically unpredictable factors determine the contingency, and thus the exact moment when a certain process is triggered off/initiated. We are well aware of that, and we respond to this impossibility by focusing on establishing, instead, which courses of events are likely, probable, or improbable.
In the sandpile model \cite{Bak1996}, to use a famous example , a crisis in the form of a large avalanche is bound to happen, but it is difficult to reliably predict when. Policy intervention on that model, if aimed to prevent large avalanches, should not act on collections of single grains, however large, but instead it should aim at defusing the basic mechanism that underlies self-organization towards criticality.

Thus, one of the key purposes of the Living Earth Simulator is to support policy and help planning intervention not based on the unattainable knowledge that a crisis is going to happen within a specific amount of time, but based on evolutionary models and new design paradigms for techno-social systems aiming at reducing self-organized paths towards critical states. Consider for example traffic breakdowns that can be caused by an overtaking maneuver of trucks. We don't know when it happens, but we know that it will happen sooner or later, when the system is in a critical state (see for instance \cite{Helbing and Tilch 2009}).

In this work, the societal forecasting paradigm will be distilled around three pillars:
 (i) the creation  and maintenance of an accurate representation of the domain under study (to answer \emph{\textsf{what-is}} questions);  (ii) the design and testing of reactions and intervention measures, which deal with issues of policy modelling and governance (to answer \emph{\textsf{what-if}} questions); (iii) the  updated representation of future scenarios of the domain (to answer\emph{\textsf{what-next}} questions) - in the words of \cite{Kauffman2011} - \emph{the adjacent possible}.
 \par The \emph{Living Earth Simulator} will enable the exploration of future scenarios at different degrees of detail, employing a variety of perspectives and methods (such as sophisticated agent-based simulations and multi-level models). The \emph{Living Earth Simulator} will require the development of interactive, decentralized, scalable computing infrastructures, coupled with an access to huge amounts of data, which will become available by integrating various data sources coming from online surveys, web and lab experiments, and from large-scale data mining (see \cite{Helbing and Balietti Visioneer} for a preliminary discussion).

 A set of software facilities, standards, and application programming interfaces aimed at the realization of global-scale socio-economic simulations and numerical models will be integrated with interactive, cooperative visualisation, analysis and information interpretation support systems. \emph{Information} about events (e.g. social unrest somewhere), trends (e.g. opinion trends), developments (e.g. consumer confidence), demographical, and other data collected on a global scale will be connected on the basis of theories, building a federation of models and translate this information into \emph{knowledge and forecastings} about global-scale socio-economics phenomena (e.g. the likelihood of a financial crisis, the expected effect of certain policies and laws, the impact on specific industries, crime rates, etc.).
 Design and implementation will fulfil  the following requirements:

\begin{enumerate}

\item Interactive exploration and understanding of big data. Novel concepts for the visualization and interaction with planetary-scale data must be created, addressing the needs of non-expert users. Through natural interaction, intuitive interfaces for exploration, planet-scale Visual Analytics, a wide range of user will be empowered to gain new insights into how society at large works.       	

\item Flexible combination and composition of different types of models (e.g. finite automata-like agents, complex agents and fluid models) across different temporal and spatial scales.         	

\item Friendly set up and  evaluation of  models and simulations. To this purpose,  problem-oriented models combined with appropriate graphical programming interfaces will be developed.  The interfaces will reduce the complexity of a explicit choice of a concrete simulation algorithm, the data exchange between models and efficient execution (including parallel and cloud computing if needed).

\item Usability. The \emph{Living Earth Simulator} will be evaluated and improved through participatory dialogue sessions, where selected users from industry, policy-making and civil society will be able to provide input through validated scientific participatory methods.

\end{enumerate}
The LES  will be the central component of the FuturICT platform, which includes the Planetary Nervous System (PNS) \cite{N3} and the Global Participatory Platform (GPP) \cite{N45}. The LES, PNS and GPP will interact, exchange data and models, and collaborate to create a truly unique basis of collective awareness.
The PNS will support the LES with the vital resource of data access, critical steps of simulation as calibration, validation and measurements of initial and boundary conditions, development of new approaches to data and knowledge mining.
The LES will be designed to be publicly accessible through the Global Participatory Platform (GPP) \cite{N45} and to have the FuturICT Exploratories as privileged interlocutors  \cite{N7,N819,N9,N10}.

The manuscript is organized as follows.
Section 2 is aimed at clarifying the main goals of the LES. This description is made more concrete by a few exemplar case studies.  Section 3 illustrates the architecture and general components of the LES. The related state of the art is described in Section 4. In Section 5 the challenges that the LES wants to address are presented. In Section 6, we will discuss the foreseeable changes that such a revolutionary artefact would have on society, that is, its impact.

\section{Setting the goals of the Living Earth Simulator}
 \label{Setting the Exploratory's goals}

Besides the scientific objectives mentioned in the introduction, the LES is driven by social goals or values consistent with human priorities and ethics principles overarched by the cultural heredity. Needless to say, these may conflict with one another: for example, open data archives may be found to contrast privacy goals, transparent information might contrast with avoiding public panic.
The \emph{Living Earth Simulator} will allow, for example, to model groups of individuals (social agents) or entire populations in spatially extended systems, and their interaction from the local scale of activities up to the global scale of mobility and transportation flows. For this purpose, the LES will integrate those mathematical and statistical modeling techniques, which have evolved from simple compartmental models into structured approaches, where the heterogeneities and details of the population and system under study are becoming increasingly important features.

Techno-socio-economic-environmental problems require computational methods and models that process Big Data, meant as large amounts of typically heterogeneous data at multiple temporal and spatial scales.

Computational models  can be seen as exploratory instruments with different purposes:
\begin{enumerate}[(a)]
\item to yield quantitative information in different scenarios.
\item to help design trajectories of development of the problems and phenomena under observation.
\item to help identify critical events in those trajectories and critical correlating variables, provided an active exchange of information with data from the Exploratories and suitable software infrastructures enabling a robust and efficient use of computing facilities.

\end{enumerate}

Several activities and tasks, data-collection and integration, and data processing aim at obtaining and maintaining a (set of) representation(s) as accurate and complete as possible of the phenomena under observation and their interdependencies. All these tasks will be implemented at the level of the Planetary Nervous System, which will offer configurable measurements of certain techno-socio-economic activities, based on data from the web (news, blogs, tweets, search volumes, ...), sensor networks, smartphone apps, web experiments, multi-player online games. The ultimate achievement being global measurements in real time (reality mining), see \cite{N3}.

\begin{figure}
\center
\resizebox{1.0\columnwidth}{!}{\includegraphics{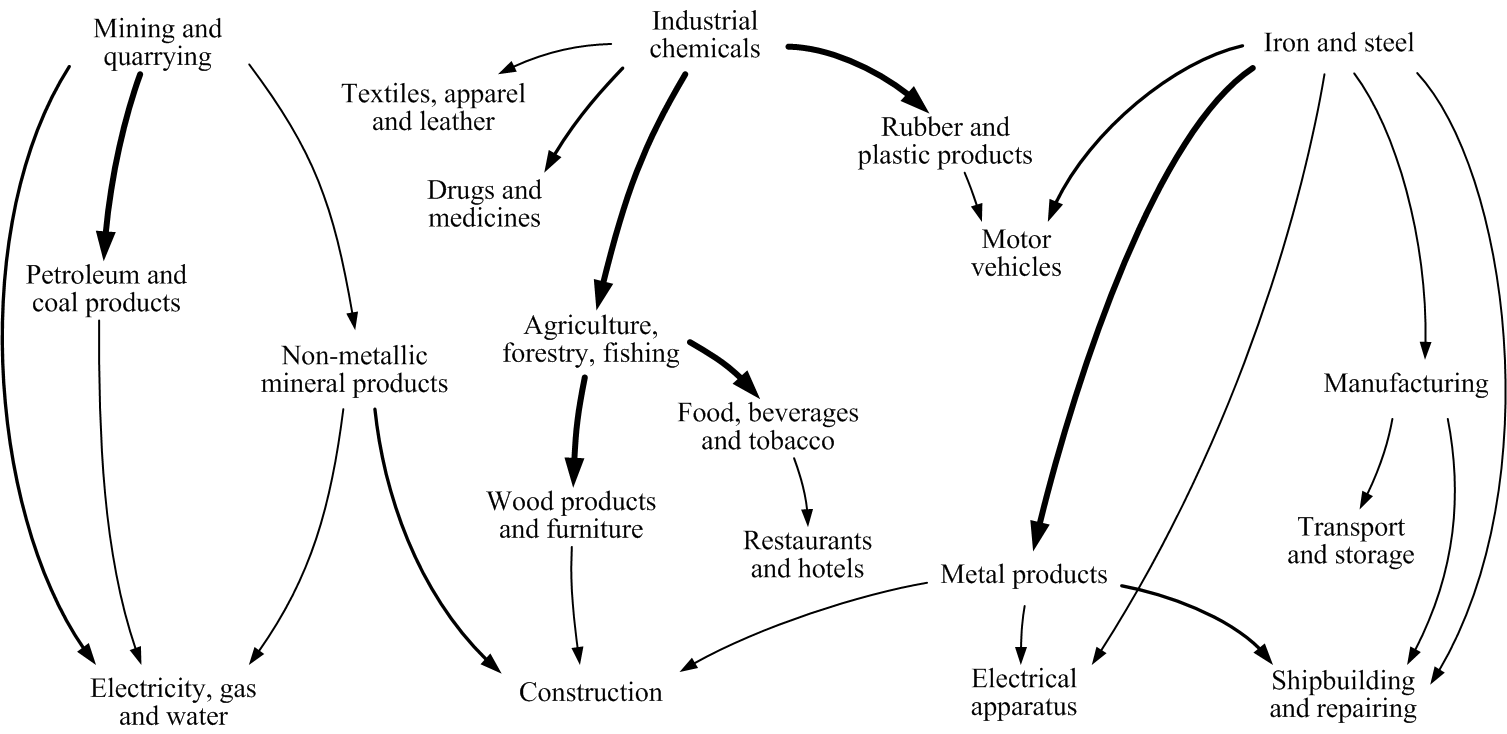} }
\caption{Causality networks allow one to anticipate likely courses of events and to prepare for possible cascading effects, for example the likely impact of supply chain disruptions (from Ref. \cite{HelbingWitt}). See \cite{xyz} for more case studies.}
\label{Causal}       
\end{figure}

In short, the \emph{\textsf{what-next}} component will provide efficient extrapolations from real-world situation , producing warnings of critical developments. The determination of causal interdependencies allows one to come up with causality networks, which facilitate to identify likely or possible next events. A concrete example of  causal interdependencies and cascade effects  flow is provided in Fig.\ref{Causal} (further studies can be found in \cite{xyz}).

\begin{figure}
\center
\resizebox{0.90\columnwidth}{!}{\includegraphics{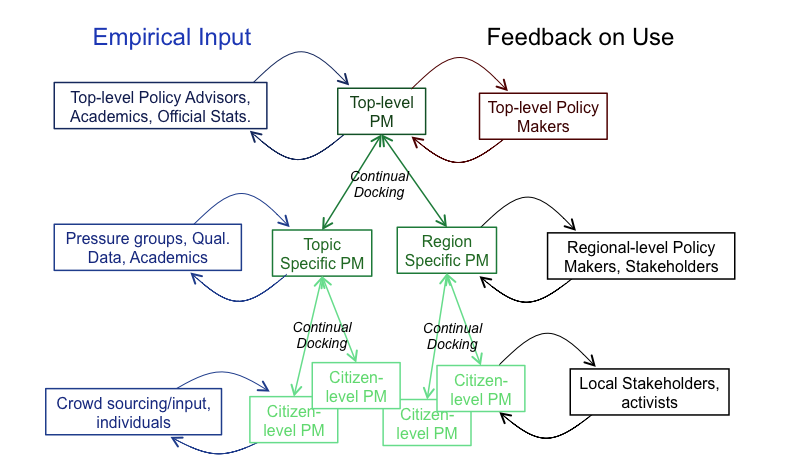} }
\caption{A Multi-level vision of Participatory Policy Modelling (PPM).}
\label{policyMod}       
\end{figure}

As discussed before cascading failures in inherent unstable systems can be avoided only by structural change. The LES will apply simulation techniques to discover the effects of such structural change in systems characterized by complex interconnections, emergent and reflexive or immergent behaviour, and nonlinear response, possibly resulting in self-organized criticality. One of the main applications of the \emph{Global Participatory Platform} is the study of possible changes in social systems, to support decision makers and stakeholders.
Parameters, initial and institutional conditions, interaction network structures and interaction rules may be changed (mechanism and systems design), and one may study what are the likely changes in the system behavior (e.g. is the system more resilient, or sensitive to perturbations or parameter changes; does it evolve into an optimal state, or does it get trapped in a suboptimal one, etc.).

The Participatory Policy Modelling (PPM) builds on LES module and generates a tighter loop involving stakeholders (see Fig.\ref{policyMod}): from providing data to specifying and criticizing the model, to determining development goals. The stakeholders' involvement comes from the relevance of their goals and their responsibility. They are presented with a wider set of alternatives, allowing them to quickly see a range of possible results.The decision-makers will have the last word in the choices, feel involved and not onerous.
\par
\begin{figure}
\center
\resizebox{0.8\columnwidth}{!}{\includegraphics{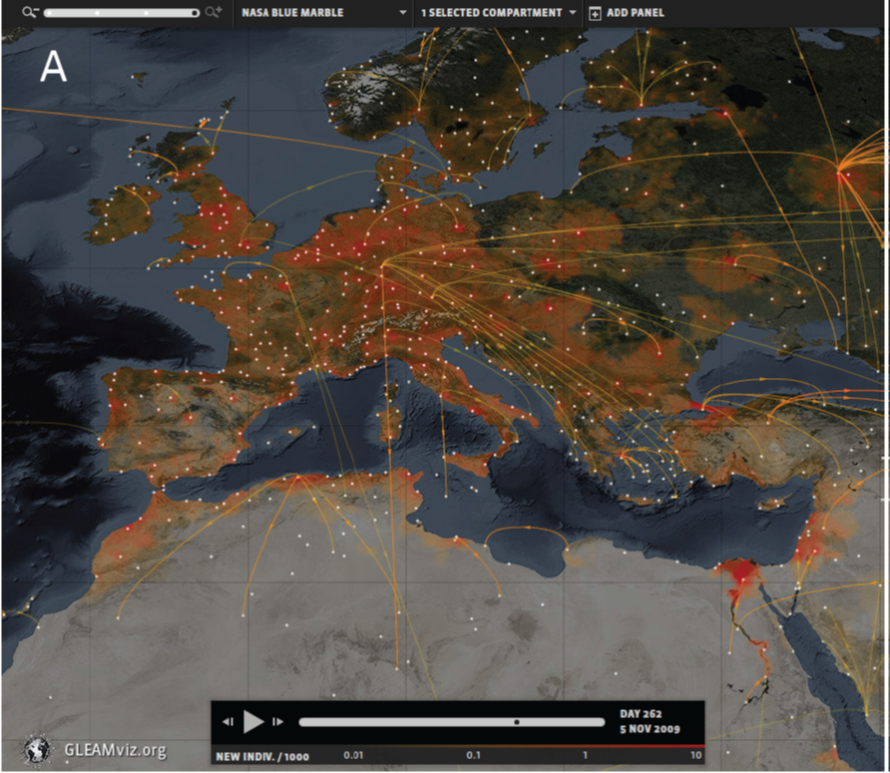} }
\caption{Color plot of the density of infected individuals on November the 5$^{th}$ in Europe without use of antivirals  \cite{Gleamviz1,Gleamviz2}.}
\label{Gleamviza}       
\end{figure}
\begin{figure}
\center
\resizebox{0.8\columnwidth}{!}{\includegraphics{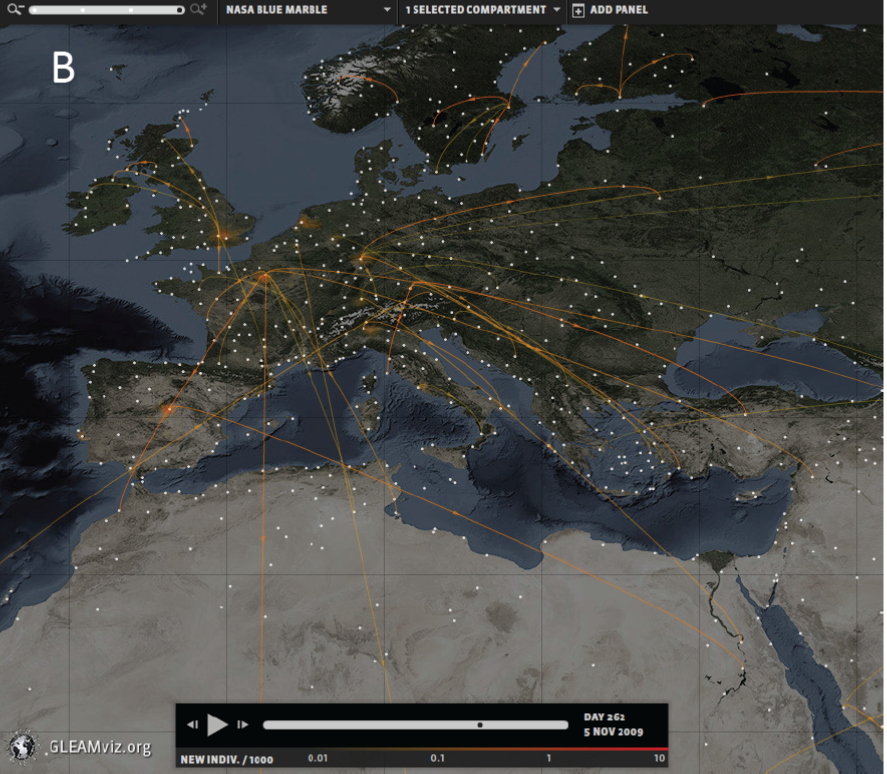} }
\caption{Color plot of the density of infected individuals on November the 5$^{th}$ in Europe  in the case of a massive use of antivirals (treatment of 30\% of the symptomatic cases) \cite{Gleamviz1,Gleamviz2}.}
\label{Gleamvizb}       
\end{figure}
\begin{figure}
\center
\resizebox{0.7\columnwidth}{!}{\includegraphics{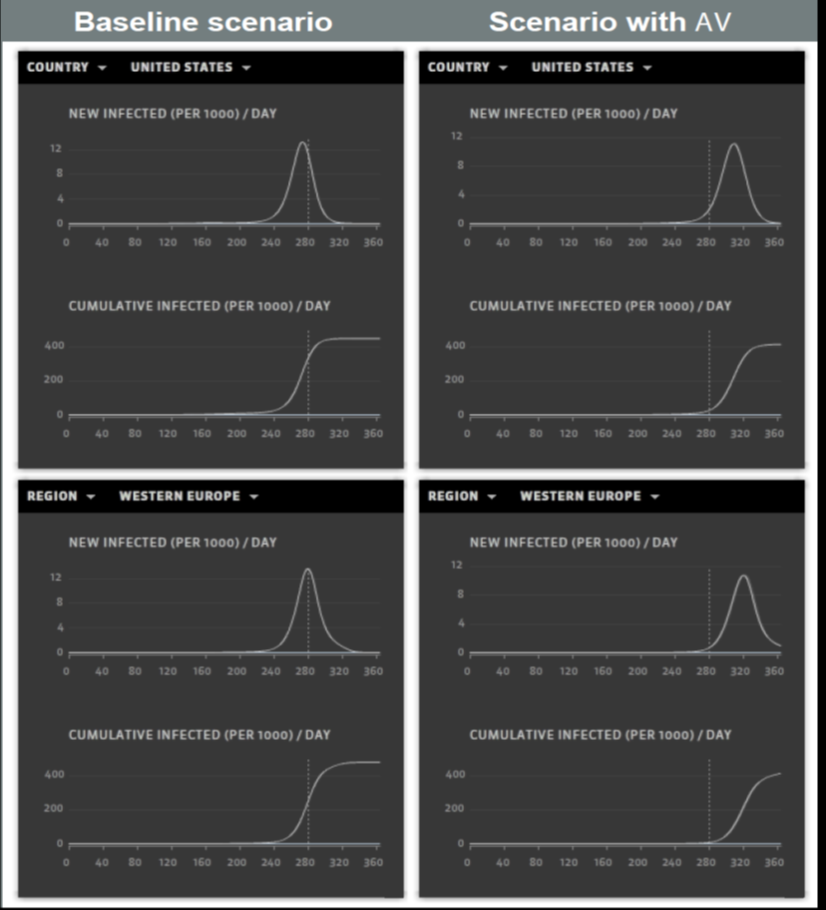} }
\caption{Scenario comparison of the timing of the peak of a pandemic event similar to the H1N12009 pandemic with or without the massive use of antivirals (AV). Simulated incidence profiles for North America and Western Europe in the baseline case (left panels) and in the AV treatment scenario for 30\% of the symptomatic cases (right panels). The plots are extracted from the GLEaMviz tool visualization.
In the upper plots of each pair the curves and shaded areas correspond to the median and 95\% reference range of 100 stochastic runs, respectively. The lower curves show the cumulative size of the infection. The dashed vertical line marks the same date for each scenario, clearly showing the shift in the epidemic spreading due to the AV treatment \cite{Gleamviz1,Gleamviz2}.
}
\label{Scenario}       
\end{figure}

  Finally, we present a case study related to the modelling and possible multiple scenarios  of the timing of the peak of a pandemic event similar to the H1N12009 pandemic with or without the massive use of antivirals (AV) obtained by the Gleamviz project (\url{www.gleamviz.org})  (see the box with the description of the Gleamviz Platform). By comparing the plots in Figs. \ref{Gleamviza},\ref{Gleamvizb},\ref{Scenario} one can readily observe the difference of epidemic timeline, with the antiviral scenario showing a very low density of infected individuals as the epidemic is delayed over several weeks.
  \subsection{Theory-building}
 \label{Theory-building}

Brut-force data mining and machine learning based on Big Data cannot solve everything. In particular it usually cannot provide an explanatory understanding. However data analytics can be augmented by theories. Therefore theory building is a transversal precondition for the activity of the Living Earth Simulator, determining its performance and interconnections. By operating at the scientific frontiers of socio-technical systems, the LES aims at investigating:

\begin{enumerate}[(a)]
\item  social problems and their interdependencies, which often requires an earlier investment in identifying the hinges among phenomena and domains addressed.
\item  challenging aspects of the dynamics of complex social systems, in particular multilevel and multidirectional processes, which still are insufficiently understood.
\item fundamental hinges, which need radically new approaches and methodologies: \cite{Liu and Tsui 2006,Conte and Paolucci 2011,Hedstrom and Ylikoski 2010}, complex nonlinear dynamics, multi-directional dynamics, multilevel and inter-level interplay  \cite{N27,N14}.
\end{enumerate}
\bigskip
\bigskip
\bigskip
\definecolor{shadecolor}{rgb}{0.941,1,0.941}
\begin{mdframed}[backgroundcolor=shadecolor]
\small
\begin{center}
\textbf{The GLEAMVIZ Platform}\end{center}\par
We have probably all experienced seasonal flu or witnessed the disruption caused by other infectious diseases.
While devastating pandemics have so far been rare, today's high population densities and intense mobility, increasingly threaten to push epidemics to pandemic proportions.
The costs of such pandemics can be immense: countless fatalities, untold physical and emotional pain, soaring healthcare costs, and supply chain disruptions.\par
Now, with Gleam we can analyse how infections may spread globally - and assess the best ways to minimise their impact, by combining sophisticated epidemic models with real-world data.
\begin{center}
\resizebox{1.0\columnwidth}{!}{\includegraphics{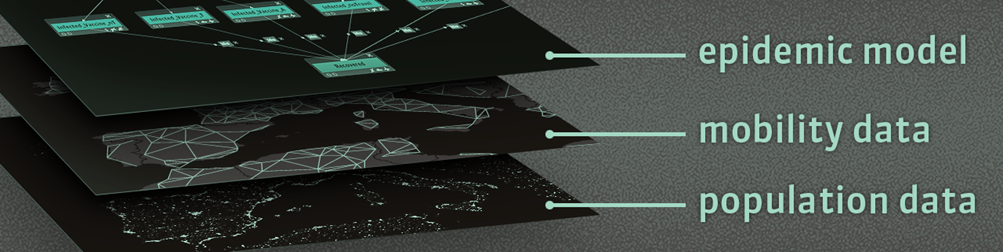} }
\label{Gleamvizc}       
\end{center}
Gleam produces accurate simulations of the global spread of infectious diseases by integrating three layers.
The first layer looks at people and their geographic distribution with respect to major transportation hubs.
The second layer adds data on the mobility of the people, how they commute, and travel around the globe.
The third and final layer adds the epidemic model which can define complex diseases scenarios and response strategies such as vaccination campaigns or emergency travel restriction.
Combining these three layers, Gleam simulates epidemic spread at an unprecedented world-wide scale. The resulting forecasts and scenario analyses help inform governments and health agencies on how best to counter pandemic threats.
Visit \url{www.gleamviz.org} to learn more about Gleamviz.
\end{mdframed}
\bigskip

These and other issues require the use of preliminary theoretical models, often based on combined qualitative-quantitative analyses and cross-methodological research, in which survey data, findings from pivotal experiments, observations and agent-based simulation studies are compared and integrated \cite{Helbing Science and Culture 2010}. If we want to obtain significant scientific advances in the understanding of social behavior, and find out what makes society different and unique, we cannot lean only on the hope that sheer data availability will make it happen. Instead, in a true cross-disciplinary approach, we will need to operate starting from theory and returning to it, because it is only the explicit statement of theories that will allow the closure of the circle between minds and society.

\subsection{Visualization and gamification}
\label{Visualization and gamification}
Visualisation and visual analytics components will play a vital role in the LES, as each of them will be confronted with massive amounts of complex data sets, data streams, ambiguous and uncertain data, data sets with missing entries, and the relations that knit them together in the shape of theories and models. Each Exploratory will help domain experts (in finance, epidemiology, organised crime, environment and many others) to make sense of data, to build simulations, to adapt old models and create new ones based on new knowledge, and to monitor various developments without getting overloaded with too much data and information.

The complexity of the data, coming both from sensors and from simulations, creates the need for powerful methods and algorithms to interactively visualize, mine and extract data, information, and models. Visual analytics approaches provide solutions as well as general visualisation techniques, virtualisation and gamification, allowing us to more readily analyze and map human processes. The power of gamification lies in its pervasiveness and in its close relation to human processes and behaviour \cite{de Freitas and Liarokapis 2011,Knight et al. 2010}. The unique properties of games allow large and complex data flows to be created and presented in relatively simple ways for users. Similarly, virtual worlds and activities are becoming more and more important in modern society and culture: one solution to planning when many stakeholders are involved is to simulate, model and gamify the complex interactions of people, traffic, transportation, mobility, energy.

\section{Setting the architecture of the Living Earth Simulator}

 \label{Architecture}

The main task of the Living Earth Simulator is to enable scientists in setting goals, building models, validating and exploring these models.

This participatory process inevitably implies a different role of the modelers as the exploratory and decision power will more directly shift to people rather than some representatives: this implies a multi-level vision of PPM, as shown in Figure \ref{policyMod}.

Building such a Living Earth Simulator involves addressing grand challenges in a number of different research areas such as computational sciences (e.g., various forms of simulation), complexity science (e.g., modeling of dynamic and interacting networks), computer science (e.g., model management, machine learning, and large-scale data analytics), and high performance computing (e.g., deployment and energy-efficient operation of data centers). These challenges will be spelled out in more detail in Section \ref{Challenges}.

\begin{figure}
\center
\resizebox{0.90\columnwidth}{!}{\includegraphics{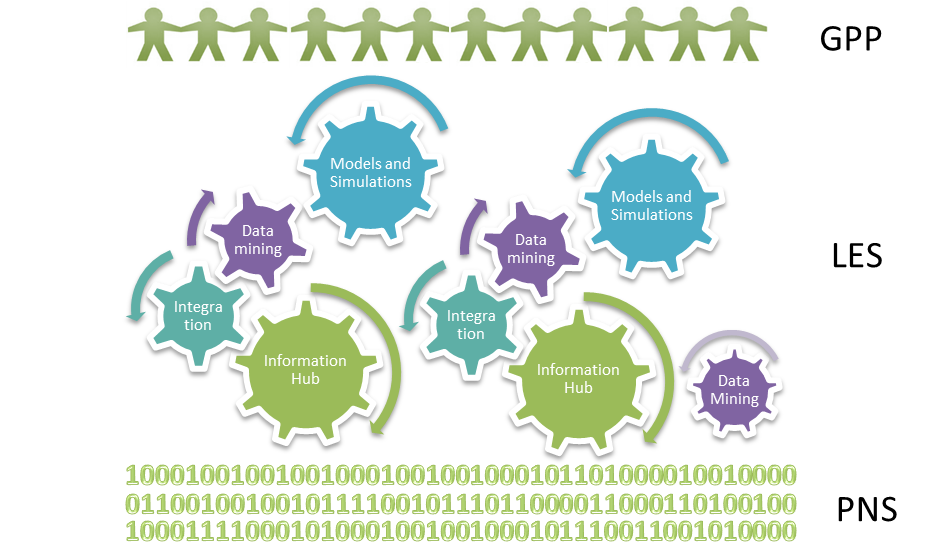} }

\caption{Building Blocks and Composition of the Living Earth Simulator.}
\label{BuildingBlocks}       
\end{figure}

This section gives an overview of the main technical building blocks of the Living Earth Simulator and how they can be composed in order to address \emph{\textsf{what-next}} and \emph{\textsf{what-if}} questions. Figure \ref{BuildingBlocks} depicts these building blocks and how they could interact in one particular instantiation of the Living Earth Simulator.

The users of a Living Earth Simulator are scientists, policy makers, market participants of a particular domain (e.g., financial markets), or, through the Global Participatory Platform, the general public. While Exploratories are specifically designed for scientists, we envision that the Living Earth Simulator, and the  analysis that it will produce, will be configured to serve a broader audience, reaching up to the general public through a suitable human-computer interface. In the current state of technology - that however is bound to change quickly during the project lifetime - we envisage those as a series of customizes \emph{Apps}. In particular, the grand challenges addressed in Section \ref{Challenges} can be applied in more general contexts.

As in every other ICT system, LES users interact with (software) applications that provide custom services. For instance, an Exploratory for a specific scientific domain (e.g., crime) will provide a specific set of applications and services to support scientists in that domain. The beauty of the Living Earth Simulator is that it provides a set of generic and powerful building blocks that address the most critical challenges in order to develop such applications. Specifically, these building blocks address the following recurring tasks:

\begin{itemize}

\item \emph{Simulations}: Computational methods and models, grounded on large amounts of data, over multiple description levels, for the analysis, simulation and exploration of complex systems (e.g., socio-economic systems). These will be addressed in sections \ref{Simulation-a}, \ref{Simulation-sda}, \ref{Simulation-ch}.

\item 	\emph{Data Management and Integration:} An infrastructure to collect, select, and integrate large volumes of heterogeneous data from possibly millions of data sources. This will be realized in cooperation with the PNS \cite{N3}.

\item \emph{Statistical Inference, Data Mining and Validation}: Machine learning techniques to analyze massive amounts of uncertain data. These techniques include state estimation to pervade and validate simulations and other applications. We will discuss them in sections \ref{Statistical Inference, Data Mining and Validation les}, \ref{Statistical Inference, Data Mining and Validation sda}, \ref{Statistical Inference, Data Mining and Validation ch}.

\item \emph{Visualization and Visual Analytics}: Tools to visualize the results of simulations and data mining and powerful interfaces that empower users to navigate through the results, create models, and initiate new kinds of simulations and analyses. For a discussion of these tools and interfaces, refer to sections \ref{Visualization and Visual Analytics}, \ref{Visualization And Visual Analytics sda}, \ref{Visualization and Visual Analytics ch}.
\end{itemize}

As shown in Figure \ref{BuildingBlocks}, these building blocks can be configured in different ways and a specific instance of a Living Earth Simulator will be composed of several of such instances of building blocks. For instance, a specific \emph{\textsf{what-if}} analysis carried out by a social scientist may involve the integration of data provided from millions of mobile phones, data mining in order to extract statistics from the integrated data set (e.g., at which times and places, users particularly frequently use their mobile phones), simulations that study the behavior of people at different times and places, and finally a validation step the correlates the simulation results from data gathered from social networks such as Facebook and Foursquare. In such a \emph{scientific workflow} each building block can appear several times in many configurations. Furthermore, the results of one step can be shared by multiple workflows as there may be millions of concurrent workflows carried out. For instance,  studies the riots in the Arabian Spring can be connected to studies financial markets by the use of the same data set: analytic results from the food economy as the prices for agricultural goods may impact the analyses of both sets. The blocks metaphor used in Figure \ref{BuildingBlocks} should illustrate both the composibility of the different building blocks and the coordinated execution of these workflows in order to achieve the sharing of results and operational efficiency.

Later on in the paper, Section \ref{Challenges} describes the grand challenges in the areas of computational science, complexity science, and computer science of some of the \emph{integration, data mining, simulation, and visualization }aspects of the Living Earth Simulator. Obviously, there are also many engineering challenges; addressing these engineering tasks in detail is beyond the scope of this vision paper. In general, the plan is to re-use a great deal of the tools and infrastructure that have already been provided in the public domain (e.g., open source software libraries). In the rest of this section, we examine briefly the three main components of the architecture, namely, Simulation, Statistical Inference with Data Mining and Validation, and Visualization.

\subsection{Simulation}
 \label{Simulation-a}

Various methodologies for carrying out massive simulations are at the core of the Living Earth Simulator, each supporting different types of models. We shall illustrate them briefly in the following.

\subsubsection{Agent-based simulation and modeling}
 \label{Agent-based simulation and modeling}

In the quest for a computational approach to social sciences, a new methodology has emerged in the middle of the 1990s \cite{Squazzoni 2010}, challenging the traditional mathematical approach which was based on systems of variables and equations connecting them, and which often restricted the scientific exploration to unrealistic models and to situations of little relevance for reality \cite{Helbing and Balietti 2011}.

Such an approach, that allows to describe computationally a society of interacting and communicating agents, is unique in that it represents complexity both inside and between individuals: from agents internal processes to agents interaction. Through Agent-Based modeling and simulation, the LES aims to represent society starting from its distinctive features; that is, as the aggregate at several interacting levels of intentional, cognitive (as opposed to rational) agents, from humans as they change in time, to humanity as it changes in time, through a myriad levels of intermediate aggregation, each of which significant even in its short-livedness and insubstantiality.

This is crucial to realize plausible models of society, permitting validation of the micro-macro connection can on both faces, and allowing for the study of emergence and for generative explanations - that is, understanding a phenomenon by explicitly running the process that generates it. Such a generative explanation, however, should be theory-based, leaning on general mechanisms and not on ad-hoc ones \cite{Conte and Paolucci 2011}.

Exploiting on one hand, the unique access to Big Data available through the Planetary Nervous System; on the other hand, the semantic structure that allows federation of Big Theories and of the models that they inspire, the simulations will be supported by access to {``}ground truth''; in this way, we can imagine getting machine learning built into the simulation models so they can be continuously adjusted to match real world data and crowdsourced interpretations. This approach would help aligning models, thereby helping with problems arising from model composition. This goes beyond model validation by getting learning and continuous adjustment {``}into the loop''.

Agent-based simulation will constitute a key component in the construction of the LES; it will provide the core tools and functionalities for the elaboration of the \emph{\textsf{what-if}} scenario that go beyond the forecast of immediate consequences. Through careful integration with theoretical and modeling components, aimed at invention and evaluation of visionary policy innovation, Agent-Based simulation  will move beyond the current fragmented, artisanal state of the discipline and reclaim its role as the language of computational society, just as mathematics was defined by Galileo the language of nature.

\subsubsection{Multiscale modelling and uncertainty quantification}
 \label{Multiscale modelling and uncertainty quantification}

Independently of agent or individual modeling, the development of computational methods and models that interface effectively with large data available by the observatories at multiple temporal and spatial scales is an essential  component of FuturICT.

Computational models may be viewed as encoders of information, thus providing a low order description of large amounts of data, as well as exploration vehicles that can provide quantitative information regarding different scenarios for the development of complex socioeconomic phenomena and help identify critical situations. These exploration vehicles rely on a close integration with the Exploratories for the active exchange of information and requires the intervention of suitable software infrastructures to enable a robust and efficient use of the available computing infrastructures.

Computing in that sense requires an interface of heterogeneous, multiscale models with massive and  heterogeneous data. It is of paramount importance to address issues of uncertainty quantification in model development along with multiscale modeling \cite{Carbone1}. Issues of model validation and verification are attenuated by the complex interactions between the various model  components and by issues of fault tolerance as expected by the implementation of these models in distributed computing architectures. There has been significant experience in modeling and simulation among the different components with large supercomputing centers routinely addressing the simulation of complex phenomena such as weather and natural catastrophes \cite{Carbone2}.

In light of the above considerations, the design principles for the modeling and simulation of complex systems will be built around these principles:

\begin{itemize}

\item  Integrated models must be able to compute their goals autonomously, within a specified temporal and spatial horizons. Goal oriented, fault tolerant error estimators must be specified for each computational model and for synergetic subsets. The assimilation and output of data from the models must be integrated with available databases from diverse sources.

\item Individual models must be integrated in a seamless multi scale computational framework. Central and distributed computing facilities must be available and be able to interface each other on demand. The need for available algorithms that make use of  parallelism is critical  in selecting models for the individual components. High bandwidth access to databases and adaptive distribution of computational resources is essential. At the same time the simulations must be robust enough to cope with limited bandwidth and with reduced data transfers. A strong component of the project is the development of uncertainty quantification techniques for the individual models.

\item Systems must be able to accommodate a multitude of computational models, ranging from completely interactive to batch execution, and their uncertainties. These models may be developed at various level of sophistication commensurate with the availability of hardware infrastructures.

\end{itemize}

\subsubsection{Self-organizing Knowledge Mining}
 \label{Self-organizing Knowledge Mining}

Many still-unsolved problems in economics, ecology, sociology, and life sciences,  are \emph{ill-defined} due to:

\begin{itemize}

\item 	missing \emph{a priori} information
\item 	possessing a large number of variables, many of which are unknown and/or cannot be measured,
\item 	relying on noisy data sets,
\item 	relying on vague and fuzzy  variables.

\end{itemize}
\noindent
which thus hamper the adequate description of the inherent system relationships.

For such \emph{ill-defined} systems, conventional approaches -- based on the assumption that the  knowledge about the world can be validated through empirical means -- needs to be replaced or supplemented to better describe their variability. This notion is based on the observation that humans have an incomplete and rather vague understanding of the nature of the world but nevertheless are able to solve unexpected problems in uncertain and unbounded conditions. This leads to a major methodological challenge: since we have an incomplete knowledge about the behaviour of the complex system (or simply, since there is no holistic theory at hand), we have to make assumptions about the missing parts to fill these gaps and be able to explain, describe, model, or predict it. By so doing, however, our assumptions might affect or even determine the result. Different assumptions might get quite different results. To yield certain outcomes, one can only go through a set of appropriate assumptions, which is sort of self-imprinting in the process of knowledge. In other words, if we are forced to make subjective strong guesses on missing \emph{a priori} information,  how reliable, adequate and accurate a predictive model of the system then can be?

\emph{Self-organising Knowledge Mining} addresses problems connected to \emph{ill-defined} systems by following an inductive modelling approach \cite{Ivakhnenko 1970,Ivakhnenko 1971} to adaptive networks based on three main principles:
\begin{itemize}
\item 	self-organisation for adaptively evolving modelling without given subjective points,
\item 	external information to allow objective selection of the model of optimal complexity,
\item 	regularization of ill-posed tasks.
\end{itemize}
Self-organisation is considered in identifying connections between the network units by a learning mechanism to represent discrete items.
 For this approach, the objective is to estimate networks of relevant and sufficient size with a structure evolving during the estimation process. A process is said to undergo self-organisation if identification emerges through the system's environment.

Self-organising knowledge mining autonomously generates models by evolving, starting from the most simple one, and validating model structure and parameters from noisy observational data, including self-selection of relevant input variables from the number of potential variables considered. It self-organises optimal complex models according to the noise dispersion of the data for systematically avoiding overfitting the design data. This is a very important condition for prediction. These models are available then explicitly in form of nonlinear algebraic or difference equations, for example.

\begin{figure}
\center
\resizebox{0.90\columnwidth}{!}{\includegraphics{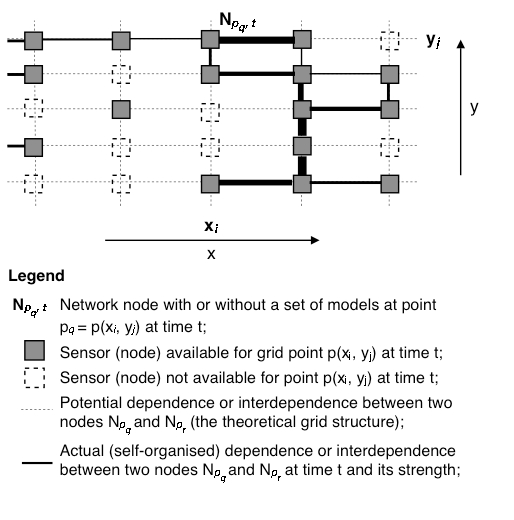} }
\caption{Self-organised sensor network model of a 2-dimensional spatio-temporal system.}
\label{SensorNetwork}       
\end{figure}

As an example, a self-organised model of a sensor network is shown schematically for a 2-dimensional grid structure in Figure \ref{SensorNetwork}. Other structures are possible, such as 3-D or $M\times M$ structures, where all nodes of a network are connected to each other potentially a priori. Each node in the network represents a subsystem of the global system described by a set of properties and characteristics. Non-homogeneity in the grid structure, which may appear and change over time, can be detected and handled by the network autonomously.

For example, a sensor network model may represents a (partial) model of the global economy. Then, each network node represents a national economy -- or, in a lower resolved level, a certain economic region. The connections between the nodes -- the interdependency between the national economies -- are self-organising from observational, noisy data. The network nodes in turn are also self-organising generating a predictive (nonlinear) dynamic system model of a national economy. This system model can be seen as another network model with self-organising network nodes (here: characteristics of a national economy).

This multi-level self-organisation allows knowledge mining for filling existing knowledge gaps in an objective way to be combined with proven a priori knowledge about the system under research, which leads to more complete, more objective, and better understanding of the behaviour of complex systems.

\subsection{Statistical Inference, Data Mining and Validation}
 \label{Statistical Inference, Data Mining and Validation les}

Simulations need to be grounded in reality. This implies the fundamental need for a statistical component of the Living Earth Simulator, which is used to

\begin{itemize}

\item inform simulations (estimating parameters etc.) through data,

 \item mine massive data produced by the simulations, and

\item validate the results of simulations (comparing \emph{\textsf{what-next}} predictions with \emph{\textsf{what-if}} analyses in hindsight, after additional data is collected).

\end{itemize}

Therefore, complementary to the simulation infrastructure detailed in Section \ref{Simulation-a}, the Living Earth Simulator will implement statistical models as well as algorithms for statistical inference and data mining.

\textbf{Informing simulations through data.} One major role of this analytical infrastructure is to provide a glue between the representation of the world  provided by the Planetary Nervous system, and the simulations carried out by the Living Earth Simulator. For example, before running simulations, one needs to set parameters (e.g., about types, frequencies and strengths of interactions between individuals and organizations), that need to best describe reality. This glue requires a unified mathematical language between these components. We will base the integration, representation and filtering of simulation data on principled foundations in statistics (using rich modern techniques such as probabilistic graphical models, nonparametric Bayesian models, etc.) and robust optimization (convex, submodular optimization, etc.). These models allow characterizing and communicating uncertainty that is fundamental in using modeling abstractions to describe the state of the world.

\textbf{Mining massive simulation data. }The statistical approach towards aggregating and analyzing simulation data also provides a natural basis for detecting trends and anomalies. By learning models of normal data, it is possible to detect significant deviations, which can be used to trigger alerts. The unprecedented scale of data analyzed will provide far more sensitive detections while at the same time providing the opportunity to decrease the rate of false alarms (since fusing data of different modalities may reduce the amount of ambiguity in the data). The approach also allows to \emph{detect common patterns }between multiple related simulations.

The tight integration between the analytical infrastructure and the simulation engine also allows us to cope with the strategic nature of the generated data. It naturally enables integration of probabilistic reasoning, e.g., by performing statistical state estimation and inference, exploring parameters according to distributions supported by the data, and strategic reasoning, by carrying out large-scale simulations of strategic agents with the estimated parameters.

\textbf{Validation of simulations.} The statistical foundations also allow to \emph{validate simulations}. While simulations cannot be used to predict the future, they allow to explore a variety of likely future scenarios. Thus, the sparse set of simulations characterizes our belief about likely futures. After new data as been collected, the common statistical language between the Planetary Nervous System and Living Earth Simulators allows to not just declare simulations as {``}correct'' or {``}incorrect'', but assign likelihoods based on the uncertainty in the \emph{\textsf{what-is}} estimate provided by the Planetary Nervous System. These estimates can then be fed back to the simulation models so that they improve with experience. We will investigate the use of techniques from reinforcement learning (which seamlessly connect to the statistical models used as a foundation for the analytical infrastructure) to close the loop between parameter estimation, simulation and validation.

\textbf{Large scale implementation. }In order to deal with the massive data produced by simulations, data mining and machine learning algorithms that are tailored for modern computing infrastructure (multicore, GPUs, cloud computing) will be developed. Current results for dealing with data streams (such as sketching, online learning, and so on) provide a basis for performing analytics and extracting information {``}on-the-fly'', without necessarily having to store all the data.

\subsection{Visualization and Visual Analytics}
 \label{Visualization and Visual Analytics}

An important success factor of the Living Earth Simulator will be the ease of combining and integrating various building blocks from the platform as part of the exploratory implementation. This aspect has to be solved by providing user interfaces, visualization, and visual analytics technologies that allow experts as well as casual users to work with the systems.

The Living Earth Simulator will utilize and create extremely large amounts of rich data (abstract, physical and semantic data) through distributed acquisition, online data feeds, and various forms of processing and simulation. Examples include data from financial markets, epidemiology, or social media simulation such as socio-economical behavior, micro-transactions, customer context awareness, large clouds of people. The data relations can be spatial, temporal, abstract and/or multidimensional. A central requirement for effective use of functional models is the ability to analyze and reason on the rich datasets created by the model. The inherent data complexity creates the need for powerful methods and algorithms to interactively visualize, mine and extract semantic model information. The LES requires a strong focus on the design and development of algorithms to visually analyze abstract and semantic data.

\subsubsection{Serious Games: the LES Virtual World(s)}
 \label{Serious Games: the LES Virtual World(s)}

Using visualization techniques, but distinguishable enough to be mentioned on its own, a central component of the Living Earth Simulator will be a virtual world that parallels the geographic and socio-economic complexity of our planet so that population density and movement, health crises, traffic patterns, conflict, natural disasters, crime, and other world-scale phenomena are coupled with the motives, desires, and actions of our planet's individuals and aggregate populations. In order to manage this enormous complexity, mixed-reality and multi-scale modelling tools will blend the real-world's complexity into the virtual environment at different scales. Likewise, complex human behaviour will be captured through gamification, bringing real-world participants into the Living Earth Simulator as living agents in the socio-economic simulations.

In everyday life, the actions of individuals and groups are deeply intertwined with and immediately affected by the geography and geometry of our world, and their analysis and simulation requires modelling the realities of the physical world, such as landscapes, city geographies, and natural resources. By combining advanced algorithms for multi-scale geometric modelling together with mixed reality concepts that blend the real-world complexity with that of the virtual environment, simulations can approach the realism and fidelity of our planet. In this large-scale virtual environment, users will not only observe the digital model of the world and navigate in it in space and time, but also augment it with their own input to edit the model, exploring multiple parallel realities and scenarios, such as shifting population masses, adding or changing roads or tunnels, expanding cities or neighbourhoods, urbanizing and deforesting, or diverting rivers. This kind of interaction is necessary to guide simulations in the virtual world and explore different options in support of informed decisions.

Gamification provides a powerful solution to move beyond the geometric reality of our Earth in order to simulate, analyse, and abstract the human processes that shape our societies. The power of gamification lies in its pervasiveness and tight coupling to human behaviour \cite{Knight et al. 2010}. The unique properties of games allow large and complex data flows to be presented in simple ways for users. The game environment of the LES Virtual World(s) will include social community tools, simulation modelling visualizations, avatars, non-player characters (NPCs), and open environments for meeting and discussions \cite{Panzoli et al. 2010}. The environment will also support user-generated content, connecting to a wide development community focused on content generation and co-creation \cite{Petridis et al. 2011}. Forums, bulletin boards, social tools and mash ups with Google maps, Facebook, Twitter, RSS feeds and web services will form core components of the collaborative-centered design. Underpinning the virtual game world will be the service-orientated architecture (SOA) allowing for interoperability with other web-based services and the portability of `avatars' across different virtual worlds.

Taken together, these components form a powerful cooperative virtual environment and mixed-reality gaming platform for scientists to explore questions  about our world's most pressing problems.

\section{State of the Art}
 \label{State of the art}

The insufficiency of current, mainly correlation-inspired approaches that try to model society is recognized by many \cite{Haas et al. 2011}, even in conjunction with the growing availability of large quantities of data - Big Data. But the integrated approach and the emphasis on possible future scenarios, provided by the LES in order to study policy implication in the wide and in the long perspective, make FuturICT unique, together with the open-access, privacy aware approach.

Corporations that are pursuing similar paths (consider for example the {``}simulating the world'' project from Microsoft or other similar initiatives listed in \cite{N1}) cannot provide the same level of openness that a publicly funded programme can. Indeed, corporation-led undertakings, as with the the Watson system from IBM, could collaborate with the LES by enabling specific directions of inquiry. But private research enterprises cannot hold the weight of the Big data and Big theories that the LES will support in an open and privacy-aware approach. Thus, no initiative exists that can actually compare to the LES, nor in the public neither in the private sector.

Of course, single components of the framework that will constitute the LES have their own state of the art which we are going to illustrate shortly in the next sections.

\subsection{Simulation}
 \label{Simulation-sda}

Simulation, and in particular agent based simulation, is currently supported by an array of technological tools and platforms, ranging in purpose from education, to rapid application development, to distributed, high performance computing (e.g. Mason) or on goal/plan based agent design (e.g., Jason). Research on reflexive simulation, that is, on simulations that recursively take into account how their results reflect into the mind of agents (as hypothesized by Castelfranchi \cite{Castelfranchi 1998}), is a promising field that is just beginning to take shape \cite{Andrighetto et al. 2007}.

Models for agent-oriented development methodologies have been developed, and reached the state of standard in the field. However, they have mostly been ignored from the social simulation community (no trace of the PASSI or ADELFE methodologies appear, for example, on the JASSS journal). The LES will be pivotal for a convergence of the different communities doing research on agent-based simulation and multi-agent engineering.

\subsection{Statistical Inference, Data Mining and Validation}
 \label{Statistical Inference, Data Mining and Validation sda}

The statistical foundations of the Living Earth Simulator build on a large body of work on data mining, machine learning and probabilistic inference \cite{Cristopher and Bishop 2006}. Rich probabilistic representations, such as graphical models \cite{Koller and Friedman 2009} provide a language and algorithms for reasoning about uncertain, structured data as arising in multi-agent simulations. These models have been successfully applied to various problems in unsupervised learning, and are naturally suited for distributed computing due to their use of message passing algorithms, e.g., \cite{Frey and Dueck 2007}.

A major challenge faced by the LES is the integration of uncertain state estimates  provided by the Planetary Nervous System with simulation models, as well as their validation in real data. While there has been a significant amount of work on learning in multi-agent systems \cite{Panait and Luke 2005}, but this work has focused on using learning to optimize multi-agent interactions, rather than mining, understanding and validating massive multi-agent simulations. Combining large scale statistical inference with massive simulations cannot be achieved by existing methods. Another major challenge is the fact that data obtained in simulations is generated from models of self-interested agents, and therefore violates classical independence assumptions widely made in statistics. While some recent work has been done on learning in adversarial / strategic domains \cite{Dalvi et al. 2004}, \cite{Bruckner and Scheffer 2011}, the problem of developing statistical methods applicable to strategic domains remains widely open.

\subsection{Visualization And Visual Analytics}
 \label{Visualization And Visual Analytics sda}

Governments, businesses, research institutes, and people themselves are collecting and hoarding large amounts of data which are presently often not utilised in the best possible way for solving the world's pressing problems. We need better and more usable solutions to extract information and ultimately knowledge from these rich data resources. Our ultimate goal as a research community is to provide visual analytics methodologies, tools, and infrastructure that will benefit society in general. The international research initiatives in the area of visual analytics including the US-American Visual Analytics Research Agenda \cite{Thomas and Cook 2005} and the European VisMaster project \cite{Keim et al. 2010} have acted as a catalyst in instigating better collaboration between leading institutes and universities working on various aspects of visual analytics.

Visual analytics is a multi-disciplinary research field, involving multiple processes and a wide variety of application areas. Within the LES, visualization will be needed both to synthesize information and derive insight from massive, dynamic, ambiguous, and often conflicting data, coming both from sensors and from simulations.

\subsubsection{Serious Games}
 \label{Serious Games-sda}

Serious games -- or educational and training games -- have become a widespread phenomenon in the wake of more pervasive game play among users of all ages. Greater broadband connectivity, faster processing speeds and accessibility of new delivery mechanisms such as mobile and casual gaming have helped to make games more pervasive in the home, at work and for leisure purposes.

Technical issues in the field have built upon advances in leading edge simulation, modelling and gaming technologies research and have often centred upon levels of fidelity and realism. Recent research directions around semantic web mash ups are looking at knowledge delivery to users through agents in 3D environments (e.g. \cite{Dunwell et al. 2010}) and reusing game content for personalisation of learning experiences. The field has also focused upon targeted application areas for use of games such as games for therapy and games for cultural heritage. In the health area, one of the earliest papers by Green and Bavelier \cite{Green and Bavelier 2003} demonstrated how video games modify attention, and later papers such as Kato and colleagues \cite{Kato 2008} have proven that game play improves behavioural outcomes in cancer treatment adherence.

Games for therapy have significant benefits for engaging younger and less motivated individuals in areas such as phobia \cite{Gerardi et al 2008,Hoffman et al 2003} and distraction \cite{Hoffman et al 2000}. In cultural heritage application areas, technical issues and user studies have followed augmented, mixed reality and interactive digital content production games, often looking at history teaching for children and augmenting museum visits with game elements. Games for changing behaviour and for supporting global altruism have become major strands in the field, e.g. the \emph{Foldit} computer game is an experiment developed by the University of Washington for folding proteins. It is a mass participation science project in the form of a multiplayer game, combining crowd sourced and distributed computing \cite{McGonigal 2011,de Freitas 2011}.

\section{Challenges}
 \label{Challenges}

Performing high quality, complex simulation experiments that consume and produce large quantity of data can require non-trivial amounts of computing power. The LES will pursue, both on the research level and on the implementation level, a federation of approaches, from High Performance Computing (HPC) and High Throughput Computing (HTC) - focusing on optimization and speed of execution -  to grid and cloud computing, focusing on involvement, open access and privacy. The combination of these approaches will provide computing capacity for the execution of large and complex models on a yet unattempted scale, thus making of the LES the unique platform where Big Problems are overcome with the integration of Big Data and Big Theories. In this section, we present the challenges that we see on the horizons of our functional components: Simulation, Statistical Inference with Data Mining and Validation, and Visualization.

\subsection{Simulation}
 \label{Simulation-ch}

\subsubsection{Agent-based simulation and modeling}
 \label{Agent-based simulation and modeling-ch}

Social systems are complex adaptive systems where agents are interacting at multiple temporal and spatial scales. Agent-based simulation addresses explicitly issues such as heterogeneity, cognition, immergence, uncertainty, social reasoning and social dependence, and is therefore cardinal to address the current challenges in the study of societal Big Problems. However, Agent-based simulation and modeling as a discipline cannot be considered ready to answer the requests that building a LES will put forward. Indeed, the very process of building a LES is likely to contribute with a renewal, or even a transformation of the field with the challenges that it will present. This is the list of those challenges that, with necessarily limited hindsight, we can foresee for the LES:

\textbf{Multiple futures}. How to apply data-mining and scenario interpretation in simulation data that may represent not just one world, but a collection of possible future worlds as simulation results? As an elaboration of the previous question, how to compare simulation result that tackle the same problem (or overlapping problems) but that are based on different, possibly incompatible theories and models?

\textbf{From individual to cognitive agents}. What is the right level of complexity within the agent? What are the problems that require agents with cognition, emotions, immergence, and in what measure? Can we define a class of problems where results obtained with simple agents will differ from those obtained with complex ones? How to model and execute complex agents on a large scale? And when that is needed, how to choose what data to retain and what data to discard?

\textbf{Massive, distributed simulations}. How to model and execute efficiently complex agents on a large scale? {``}Smart parallelization'' is strongly needed for multi-agent-based simulations in order to accommodate huge models, that are needed to model meaningful complex social and economic phenomena.
Thus, we must ensure efficiency for a simulation that involves billions of complex agents but nevertheless has to ensure results in reasonable time. To solve this problem, parallelization techniques over different and heterogeneous hardware of large, complex and diverse simulations will need to be studied.

\textbf{Large-scale verification and validation.} Simulations should be designed, implemented, thoroughly tested and debugged, but automatic tools that scale up to LES size have yet eluded the research community. The need for parallel languages, supporting an automated testing and debugging for huge simulations, possibly under multiple models alignments, is therefore a paramount objective of this research line. This will require the development of a common toolbox of a federation of tools.

\textbf{From mega models to global social awareness.} The paradigm shift that is the objective of FuturICT needs a whole new way to conceive the creation of societal future. This should include agent-based simulations that are easy to design and access. We need to connect the models with the results of simulations based on these, in an easily communicable form, and we need to recursively take into account the effects of these results when they are communicated to people. Here, the challenge for simulation is to play a pivotal role between constructs that each of them are a challenge by themselves.

\subsubsection{Multi-scale simulation}
 \label{Multi-scale simulation}

The simulation of complex techno-socio-economic systems requires the development of effective multiscale computational methods for models of complex systems as they pertain to individuals and hierarchical emerging structures, as well as the interactions among themselves and their environment. The environment can be understood in terms of natural phenomena such as weather and volcanic eruptions as well as artificial structures (here called Organizations) such as companies, cities and even social and religious beliefs. Each of these components manifests itself by the production and processing of data and the exchange of information with its environment. Modeling these systems and their interactions requires the integration of data with multiscale computational models developed not only across disciplines but also across world wide distributed data Exploratories and computer infrastructures. Novel computational tools are necessary to handle the associated communication and computational complexity induced by the interaction of the above-mentioned components. A fundamental computational difficulty lies in the heterogeneous, multiscale complexity of these systems and their representation.

We may distinguish the following computational modeling problems  as closely linked and interacting across temporal and spatial scales:

\textbf{The Self} The world is composed of individuals who act based on their cognitive state and the information they receive through their interactions with other individuals and the environment. The cognitive state of the individuals may be the result of their interaction with organizations (societal influences, religious beliefs, etc.) and the environment (natural phenomena, cities, etc.). These interactions are asymmetric in terms of their strength and effects. The computational self is characterized by its capability to alter the behavior through learning and by the desire to survive. Individuals interact with information with a relatively limited {``}bandwidth'', while individuals are among the main producers and targets of information through their behavioral patterns.

\textbf{Organization.} The interactions of individuals gives rise to collective structures that may be broadly defined as {``}institutions''. These may be represented in form of artificial structures (social groups, cities, companies, etc.). They need to be modeled at multiple levels of description that acknowledge their heterogeneous composition and prescribe interactions with individuals and the environment. Organizations are considered as having much higher bandwidth in the absorption of information than individuals. In turn, organizations often have a geographically  distributed existence and produce information and interact with the environment  in ways that have a much larger impact than that of the individuals. The modeling challenge involves the identification of the essential parameters that identify organizations. We may consider characteristics with a   scale-invariant complexity that links individuals and organizations and at the same time distinguishes their different behaviors and interaction. We need to quantify the information processing in organizations and identify the multitude of cost-functions in organizations (survival, societal influence, financial profit, etc.).

\textbf{Environment.} It is broadly defined as the set of inanimate natural and artificial phenomena. We distinguish natural phenomena such as the weather and volcanic eruptions as well as engineering products such as cars and nuclear power plants. There is a large tradition for modeling such systems in the fields of Computational Science and Engineering and this expertise will be very valuable in this project. However, existing paradigms must be reconsidered in terms of the types of the scale and reliability of the heterogeneous computer architectures (Cloud Computing vs/and Centralized Supercomputers) and the previewed intensive interaction with Data Exploratories.

\subsection{Statistical Inference, Data Mining and Validation}
 \label{Statistical Inference, Data Mining and Validation ch}

There are several fundamental challenges in developing the statistical foundations of the Living Earth Simulator, which go beyond what current methods can handle.

\textbf{Massive scale.} The scale of simulation data produced by the LES goes very far beyond what existing data mining and machine learning algorithms can handle. There may even be too much data to store, and thus it would have to be processed and aggregated in real-time while being produced.

\textbf{Cross-methodological approaches}. Simulation, serious games and statistical inference have been traditionally developed independently of each other. The Living Earth Simulator must bring these together to allow joint reasoning and analysis. This requires the development of common representations, as well as approaches for reconciling potentially conflicting predictions by the different methods.

\textbf{Rich statistical models}. It is necessary to find the unified language to close the loop between the Planetary Nervous System and the Living Earth Simulator. This language must be able to express and quantify uncertainty inherent in answering \emph{\textsf{what-is}}, \emph{\textsf{what-if}} and \emph{\textsf{what-next}} questions. It must be rich enough to express complex relations between concepts explored in the simulations.

\textbf{Principled approaches for validating simulations}. Long term simulations are naturally difficult to validate empirically. We need to characterize classes of simulation tasks where accuracy can be diagnosed within a short period of time. This requires the development of a learning theory for \emph{generalization }in the context of simulation.

\textbf{Nonstationarity and rare events.} LES needs to detect and  cope with non-stationary phenomena  (e.g., economic trends), as well as rare events (shocks, crises) for which little or no prior data may be available.

\textbf{Data from strategic agents.} The data collected and processed by the LES is partly generated by strategic agents (e.g., stock traders). Learning theories therefore need to be developed  for reasoning about strategic data.

\subsection{Visualization and Visual Analytics}
 \label{Visualization and Visual Analytics ch}

The following subsections provide the details on the challenges  for enabling successful visualization and visual analytics research within FuturICT. The challenges and recommendations highlight in particular the inherent interdisciplinary nature  of such a joint research agenda.

\subsubsection{Visual understanding of data and knowledge discovery}
 \label{Visual understanding of data and knowledge discovery}

Visual analytics and visualization are concerned with data, users, and designing a technology that enables the user to visually and interactively make sense of large and complex amounts of data in order to extract information and augment their knowledge. The visual analytics process aims at tightly coupling automated analysis and knowledge discovery methods, and interactive visual representations. Shneiderman \cite{Shneiderman 1996} proposed a guide to visually exploring data presenting it on the screen through his mantra: {``}Overview first, zoom/filter, details on demand''. However, it is difficult to create an overview visualisation of large and complex datasets without losing interesting patterns. Without an overview, zooming and filtering techniques do not help users, because they have little information of what to examine further. Daniel Keim \cite{Keim 2006} has extended the guide for visual analytics: {``}Analyze first, show the important, zoom/filter, analyze further, details on demand''. This new \emph{mantra} indicates that is not enough to retrieve and display the data through a simple visual metaphor. It is necessary to look at the value of interest and analyze the data according to it. In this way, visualization can show the most relevant aspects of the data and at the same time provide interaction models, which allow the user to get details of the data on demand, again connecting to the knowledge discovery, data mining tool or automated algorithm to support the operation.

Practitioners in visualization and visual analytics have implemented several solutions, such as in-memory databases or user-steerable algorithms. However, these are isolated attempts and not sustainable solutions in the long term. They do not provide a unified platform and the components cannot inter-operate at a larger scale. This exactly is the goal of the Living Earth Simulator and its envisioned service for the Exploratories. Visual analytics adds several scientific challenges to the interoperability of techniques both on the technical and the methodological aspects of data analysis and simulation.

As an example, a group of researchers needs a component or knowledge from outside their native field, which fulfills a specific uncommon requirement. The LES will encourage cross-domain research, because it is a leverage point to search for solutions, competencies or requirements beyond their own field. This illustrates that the LES does not only draw upon knowledge to integrate existing techniques. We expect transparent mutual requirements to have a high impact on the research in each core discipline.

Core scientific challenges include a theory for quantifying visual information going beyond the Shannon models of information theory. Based on such models, we will be able to define visual information elements and building blocks as well as visual metrics for abstract data spaces. In addition we will design methods for projection of data into meaningful subspaces that can eventually be visualized. All of such algorithms will have to consider the online nature of the model as well as the realtime data feeds making consistency of model and visualization a nontrivial issue. For effective visual data mining the user has to be able to feed relevance back into the system and to control the simulation. We will focus on user adaptive algorithms based on machine learning.

To achieve this goal, with respect to visualization and visual analytics we see the following challenges for research on the LES:

\textbf{Design Guidelines for Big Data.} One of main challenges is to utilize our existing theoretical and practical knowledge by making it readily available to designers of LES-based Exploratories, possibly in the form of design guidelines. For instance, there is a wealth of experimental results in the field of visual perception and cognition that would be of considerable benefit to interaction designers, if it were organized appropriately. For a given task, the challenge is to provide guidance on what to use (e.g., method of analysis, type of visualization), how to use it and how to decide if it was a good choice.
Another major challenge is dealing with the management of very large datasets, whether this is in terms of storage, retrieval, transmission (as with distributed databases or cloud storage), algorithm processing time, and scalability of visualizations. Data is often heterogeneous and can be of poor quality with, missing, incomplete, or erroneous values. This adds to the complexity of integrating data from many sources. In addition, data often requires transformation of some sort (e.g., scaling and mapping) or requires specialized data types, which are seldom provided by current database systems.

\textbf{Real time interaction.} Streaming data presents many challenges -- coping with very large amounts of data arriving in bursts or continuously (as with analyzing financial transactions or Internet traffic), tackling the difficulties of indexing and aggregation in real-time, identifying trends and detecting unexpected behavior when the dataset is changing dynamically. Semantic management (managing metadata) is currently not well catered for, despite the wealth of information contained in rich metadata.

\textbf{Dynamic simplification and summarizing.} Another challenging aspect is using visual analytics to simplify the models and patterns extracted by advanced data mining techniques. Existing methods are largely non-intuitive and require significant expertise. Similar efforts are required to assist users in understanding visualisation models, such as the level of abstraction (actual data or aggregated view) and visual metaphors (how the data is represented on the screen). Expert analysts require this flexibility and so do lay users, who in addition, require guidance in, for instance, choosing appropriate analysis tools and visualization methods for the task at hand. Users often wish or need to collaborate in order to share, or work cooperatively on, the data, results of analysis, visualizations and perhaps workflows. Providing the necessary distribution infrastructure as well as the user interface is a challenging task.

\textbf{Designing an open modeling platform for data management, analysis, exploration and visualization.} One goal of FuturICT is that experts from different fields of research benefit from each other's work. The LES will provide a common language for different fields related to complex systems analysis. It enables researchers to formalize and understand their mutual requirements and their results, and to understand their role and contribution within the analysis process. It is based upon the current experiences of practitioners in their native fields, but it heads towards filling the gaps at the boundaries between the different fields.

\textbf{Designing user-driven analysis and simulation techniques and methods.}The LES will allow each ICT domain to implement so-called building blocks (from a visual analytics viewpoint: information visualization methods, data mining algorithms, data mining algorithms, data management approaches) in a coherent way, leading to compatibility and interoperability across the board. Current data analysis techniques are data-centered: they read data, run to completion and write their results. Recent research has led to new paradigms for large-scale computations for modeling and data mining (grids-based systems, GPU systems, computing in the Clouds). However, these novel paradigms are all data-driven. Exploratories need user-driven methods instead. Thus, we should explore how new approaches could allow interactivity in combination with high performance. We envision several methods and strategies that need to be fit into the Living Earth Simulator and be individually validated. Validation means that they should be practical to implement and provide results with a quality and throughput similar to their non-user-driven implementations.

\textbf{Validating the building blocks across the LES.} Once the platform is specified, there will be a common ground for benchmarking various visual analytics, simulation and other building blocks that emerge across different ICT fields of research. Several European companies and institutions offer components in that direction, but their user base remains small at this level compared to other software components. Comparability through the LES will be an important driver for excellent research and development in FuturICT.

\subsubsection{Serious Games}
 \label{Serious Games-ch}

The implementation of serious gaming environments -- both multi-player and mixed reality necessitate overcoming specific challenges. While gaming platforms are currently capable of integrating different data sets and querying real time across disparate data sets, issues around real time visualisation, feedback modelling and integration of real time large scale simulations have not been tested outside the research environment. To deliver stable environments to large numbers of students, scientists and general users will require combining different gaming platforms with multiagent modelling and large scale simulations in real time and dynamically. While the tools are currently available the combination of these is still at the research stage. The main challenge therefore will be to develop a stable infrastructure and architecture that can blend inputs from mobile technologies, databases and simulation platform in a seamless interface for users. The capability to personalise this interface and to deliver content to different devices over an open source gaming platform will rely upon robust technologies and platforms, as well as a considered interface development using participatory development models. The gaming platform is envisaged as a wrapper or interface for social interactions with datasets, modelling human processes and behaviour as outputs of the LES and as data injects. The approach is to use the gaming platform as a {`}reciprocal' system, where users can co-create and deliver content and test and hypothesis through gaming and simulation. The challenges therefore will be technical: to combine the different technologies and platforms together, social: to develop human process and behavioural models and political: to develop reciprocal systems whereby real time analysis of data modelling and data inputs contributes to scenario-development, decision-making and research hypothesis formation and evaluation.

\subsubsection{Accessibility}
 \label{Accessibility}

A specific challenge is to make the massive numerical representation of scientific results, including uncertainties, comprehensive to planners, policy-makers and the public. Adapting and applying visualization techniques may significantly assist in analysing and communicating inter-linkages, complexity and scientific uncertainties. Collaborative Visual Environments facilitate the use of intuitive tools for understanding complex challenges. The high level of interactivity when, for instance, comparing scenarios, projections and models, improves the potential of making them comprehensible for decision-making. These tools assist dialogs so that identified linkages are more comprehensively addressed in local, national and international policy processes. The techniques and methods  have advanced socio-economic research by transforming complexity as a cause for inaction to multi-dimensional data integration for robust and transparent analysis.

The methodology for visualisation-assisted dialogues is based on the iterative validation of input data and use of model results in a participatory process that aims at ensuring that the involved stakeholders have a high degree of confidence in the LES.

The participatory component in the Exploratories, will build on collaborative visual environments where relevant stakeholders are consulted, and can access, reflect upon and discuss expert knowledge. They can be in three forms of usage: 1) the Arena Station where participants interact with other through desktops; 2) the Conference Arena in the form of a round-table discussion using multiple displays where the participants interactively can display e.g. data, scenarios and models; 3) The Collaborative Arena where a large group of people use interactive ICT-based visualisation techniques to interact similar to the Conference arena, but on a larger screen, such as dome display environments. Analytical tools that assist understanding of linkages across sectors and scales will be designed to enable visualization of time-dependent, multi-parameter data sets.

The LES will carefully apply socially-enhanced methodologies for domain-agnostic knowledge extraction from globally distributed repositories, with a mix of automated processing and expert contributions. Data from different sources will be accessed, analyzed by geographically distributed experts discovered on demand and visualized considering the different (maybe advanced and unconventional) user devices. This approach will enable online users and expert communities to collaboratively (re-)evolve rich analytical or operational reports and dashboards produced by knowledge flows in real-time from a myriad of devices upon the vast knowledge made available through the ecosystem and to create new knowledge within this ecosystem.

Visualization techniques in combination with participatory methodologies can be appropriated to support decision-making and stakeholder interaction. As previous research on model assisted stakeholder dialogues indicates, the outcomes of the use of  for example, a quantitative model cannot be separated from the participatory process itself. This raises the importance of embedding the use of visualization techniques in a real world process context and also highlights process design issues as crucial. Thus, the identification and involvement of process participants, the sequencing of meeting contents and efforts to ensure that also dimensions not captured by quantitative data sets are included in the sharing of knowledge's are issues that require careful attention.  This involves research on social and institutional learning and development and testing of intermediary concepts and objects that enhance boundary spanning visualizations and uses different types of data.

\subsubsection{Usability}
 \label{Usability}

The use of knowledge in decision-making has proven to be a complex web of interactions. The traditional image of {`}truth speaks to power' has been rebuked by views that emphasize the mutual interplay or co-production of science \cite{Sarewitz and Pielke 2007}, with the aim of making science more {``}socially robust'' \cite{Nowotny et al 2001} through direct engagement with the societal context. The participatory dialogs facilitated through LES and the Exploratories entails modelling \emph{with }people, as a complement to the agent based modelling, which implies modelling \emph{of }people's behaviour or attitudes \cite{Pahl-Wostl 2002}. Knowledge co-produced through deliberative dialogs between researchers, policy makers and other stakeholders is expected to provide new perspectives, contextualize findings, and probe assumptions (e.g. \cite{Wilsdon and Willis 2004}).

Participatory research is particularly relevant in areas of high uncertainty or high stakes (e.g. climate change, biotechnology). The different {``}knowledge-abilities'' of lay people and stakeholder groups are expected to augment the scope and quality of scientific risk assessment as well as the legitimacy of potential solutions (cf. \cite{Funtowicz and Ravetz 1993,Jonsson et al 2009}). By engaging in deliberation on uncertain and ambiguous aspects of problems facing society, participatory methods can increase society's ability to deal with stochastic and unpredictable challenges \cite{Nowotny et al 2001}. If social actors directly affected by research results are invited to validate the assumptions made in the various steps of a research process, it is assumed that they will gain trust in the findings \cite{Jonsson et al 2009}.

There are many challenges related to system usability and process understanding. For users to have confidence in the data they should be aware (or be able to discover) where the data comes from, and also what transformations have been applied on its way through the process pipeline (e.g., data cleansing, analysis and visualization). Furthermore, a clear understanding of the uncertainties in the data and results of the analysis can help minimize cognitive and perceptual biases, which without attention can significantly affect the interpretation of the results.

The degree of interactivity is important for all users. Rapid feedback is critical in visual interfaces and this presents challenges to many of the domains associated with visual analytics. Evaluating visual analytics applications is particularly difficult due to the complexity of human interaction with multiple processes (e.g., analysis and visualization). The question of how to classify success or decide what is a good solution, is problematic when dealing with exploratory tasks, which are typically ill-defined or open-ended.

Beyond the core visualization issues, the research on usability and HCI will be focused on two challenges: situated interaction with visualizations, and designing to support sensemaking.

The study of situated interactions with future technologies poses many challenges: how can large bodies of interrelated data be presented to people via devices with different form factors, via different modalities, such that an individual can keep track of where they were across transitions between one device (with one set of display and interaction capabilities) and another? Various approaches to this have been investigated (e.g. \cite{Bandelloni et al  2004,Chang and Li 2011}), but none of these has tackled the issue of scalability, and the fact that devices with significantly different form factors will support significantly different information presentation and interaction. Research within this strand will employ innovative evaluation methods to establish user needs and expectations, and prototype novel interfaces across a range of state-of-the-art (as well as established) interactive devices. It will work with representative user populations, involving studies in the home as well as the workplace and other (semi-)public spaces.

It doesn't matter just that people can access information, but that information is meaningful to them -- i.e. that they can make sense of it. Sensemaking is usually relative to some pre-existing information goal, whether well-articulated or not. Most research to date on sensemaking has focused on professional sensemaking, often working with large bodies of data (e.g. \cite{Attfield et al 2011}); relatively little research has been done on how lay people might make sense of the same data, and little is known about under what circumstances members of the public might engage with a resources such as the LES, or how the various potential uses (e.g. in education or to support national debate) can be effectively supported. Research within this strand will seek to better understand what might motivate people to engage with the LES, as providers and users of data, how they make sense of information, and how they use it to make decisions, to build their understanding of the social world of which they are a part, and to engage better as citizens of the world.

\subsection{Computing Infrastructure}
 \label{Computing Infrastructure}

 The scale of the questions to be answered by the FuturICT platform requires the usage of resources of comparable  scale, particularly for what concern the computational infrastructure. Existing tools and infrastructures will be expanded in order to tackle the enlarged scales.

In the {High Performance Computing} regime, the PRACE (\url{http://www.prace-ri.eu/}) Research Infrastructure (PRACE RI) enables European scientific discovery and engineering research across all disciplines for the benefit of society. World class computing and data management resources as well as services open to all European public research projects are provided through the PRACE RI. These resources can be used for providing state of the art results and their usage is optimal in cases of highly scalable and efficient parallel applications. On such types of resources porting of applications and optimization studies are non trivial operations that are nonetheless essential for making efficient usage of the offered resources.

A more decentralized and {distributed computing} and storage infrastructure is offered through EGI.eu  (\url{http://www.egi.eu/
}), which maintains the pan-European Grid Infrastructure (EGI) in collaboration with National Grid Initiatives (NGIs) and European International Research Organizations (EIROs). Bringing together and combining resources spanning the globe is an excellent infrastructure for loosely coupled parallel applications for processing data. The complex science research community has been using computing and storage resources through the Complexity Science Virtual Organization (vo.complex-systems.eu). In this context of we plan to further build upon this Virtual Organization infrastructure by provisioning more resources and developing state of the art services for its members.

{Cloud computing} has recently emerged  as a natural evolution towards the provision of more distributed cost efficient computing resources. Several computational methods that are based on the Map Reduce model have been optimally designed for such infrastructures and these are now starting to gain trend in the research community. We plan to develop and provide back ends for performing Map Reduce operations on top of Cloud based resources focusing on interactivity and analysis of \emph{{Big Data}}. Towards this direction, we will be focusing our activities on the following areas:
\begin{itemize}

\item 	distributed storage and transparent access of highly complex data sets

\item 	scientific data management, search and filtering

\item 	high-level data-analysis (feature extraction, statistics, time-series) to investigate and describe complex data-relationships

\item 	portal solutions for convenient access and graphical interaction.

\end{itemize}

\subsection{Cross Methodological Aspects}
 \label{Cross Methodological Aspects}

\subsubsection{Social Dynamics}
 \label{Social Dynamics}

To trace interesting dynamics of complex social systems it is necessary to investigate fundamental cognitive-behavioral and cultural properties: cognitive biases drive the behavior of financial systems, uncertainty is the main cause of financial instability, and the very nature of {``}credits'' brings about the underlying cognitive dimensions of credibility, expectation, reputation \cite{Paolucci and Conte 2009}, overconfidence \cite{Akerlof and Shiller 2009} and trust \cite{Castelfranchi and Falcone 2010}. Epidemics are accounted for in terms of social distance \cite{PerraVespignani2011}, which is a highly culture-dependent variable. Criminal systems of the extortion type fit the culture of honor \cite{La Spina 2008}, which they contributed to create; but the primeval soup, the cultural grounds of corruption, after more than a decade of statistical analyses and comparative surveys, is still unclear \cite{Della Porta 2004}. It is not always clear when you need to investigate the underlying mechanisms, and when the statistical analysis is sufficient. Which social phenomena may be accounted for in terms of cascading effects? Which interconnections and interdependencies should we expect to find, and therefore should pay attention to, in the spread of social conflicts? When is social unrest more likely to result in open conflicts? On the contrary, where should we put our sensors in the attempt to predict democratization? Why do we expect that the new media would support people in getting rid of dictatorships instead of being exploited for surveillance and repression?

Finally, human attention is a scarce resource. Another important challenge for developments of the theory-building component of Exploratories is grounding recommendations on the most relevant observations (trends, anomalies, criticalities). These should be selected to fit cognitive biases and attentional boundaries of domain experts (financial analysts, etc.), who are expected to understand and act on those recommendations. Interactive analyses and participated decision-making concerning activities in the activities of different components -- especially while recommending simulations to be carried out in the \emph{\textsf{what-if}} component - should be supported by the Exploratory collaborative infrastructure.

\subsubsection{Cognitive Behavioral aspects}
 \label{Cognitive Behavioral aspects}

Although many basic conceptual questions remain unresolved, the major challenge in the development of models able to capture the behavior of large-scale techno-social systems is their sensitivity and dependence on social adaptive behavior. In the absence of stress, techno-social system tend to reach some kind of stationary state is reached in which the feedback between the social behavior and the environment determines the details of how the dynamical process of interest play out. But social behaviors react, adapt and define new way of interacting as the dynamics of the system evolves. Contrary to what happens in physical systems, the global evolution of the system and the knowledge of it are part of the system dynamic. While some of the above issues may find a partial solution by improving the accuracy and reliability of models, it is clear that the social adaptation to predictions face us with new methodological and ethical problems.

Addressing these problems involves tackling three major scientific challenges. The first is the gathering of large-scale data on information spread and social reactions. This is not presently out of reach, via large-scale mobile communication databases (mobile telephone, twitter logs, social web tools) operating at the moment of specific disaster or crisis events. Second is the formulation of formal models that make it possible to quantify the adaptation, changes and reactions of individuals as a function of the dynamical processes occurring in the system. The third challenge concerns the deployment of monitoring infrastructures capable of informing computational models in real time.

\section{Impact}
 \label{Impact}

Governance is no longer confined to the territorial boundaries of the nation state, but increasingly needed across geographical sites (e.g. local, regional, national and transnational). In this {`}polycentric' governance landscape \cite{Ostrom 2009}, we have seen the rise of {`}softer' and less hierarchical forms of steering that rest upon collaboration among government, business and civil society actors. A general assumption is that such {`}transnationalisation' of governance will lead to more effective and legitimate policies \cite{Rosenau and Cziempel 1992,Cerny 2010}. The Living Earth Simulator is designed to meet the challenges of this new broader governance landscape, by providing a platform for visualization of linkages between geographical scales, and interconnectedness between governance arrangements at different management levels.

With societies becoming more interconnected and systems developing dependencies, new challenges are emerging that recast the role of social interaction for problem solving. The interconnected nature of our systems  is leading in particular to self-organised criticality in many areas - economy, finance, institutions, politics, making our systems vulnerable to collapse and reorganisation cycles that are rapid and hard to predict and therefore difficult to design. The design of the LES  aims at addressing this issue at its heart by utilising the new technologies and structures in socially constructed and driven contexts thus impacting society and economy. Specific areas of interventions are smart cities and  energy systems \cite{N16,N17}, for instance, human consumption and natural resources are at loggerheads, politics and socio-economic factors pervade everyday life and human over population, crime and poverty create difficult problems for planning, transport and smooth operations.

The Living Earth Simulator can bring together social science research into an experimental setting, creating a rich ground for experiments with virtual archaeology, clustered scenario building, forecasting data flows and systemic behaviours and inter-service training patterns. This rich ground for scientific and social breakthroughs will provide for innovative and novel lines of research in anthropology, geographical data and population forecasting; other research benefits will include consideration of virtualization and gamification for supporting collaborative research, hypothesis formation and virtual communities as research tools and experiments, for testing frameworks and tools and for supporting validation of research hypothesis.

\section{Acknowledgements}
The GLEAMviz project and Alex Vespignani are gratefully acknowledged for providing Figs. 3-5. Frank Lemke is acknowledged for the text in Section 3.1.3. \par
The publication of this work was partially supported by the European Union's Seventh Framework Programme (FP7/2007-2013) under grant agreement no.284709, a Coordination and Support Action in the Information and Communication Technologies activity area (`FuturICT' FET Flagship Pilot Project). We are grateful to the anonymous reviewers for the insightful comments.

\end{document}